\documentstyle [epsfig,epsf,rotate,]{mn}
\newcommand{\xp}[2]{{\partial #1 \over \partial #2}}
\newcommand{\xpp}[2]{{\partial ^2 #1 \over \partial #2 ^2}}

\title{$N$-body simulations using customised potential-density pair basis sets}
\author[M.J.W. Brown and  J.C.B. Papaloizou]
  {M.J.W. Brown and  J.C.B. Papaloizou \\
   Astronomy Unit, School of Mathematical Sciences, 
   Queen Mary \& Westfield College,
   Mile End Road, London E1 4NS}
\begin{document}
\maketitle
%
\begin{abstract}
  Potential-density pair basis sets can be
  used for highly efficient $N$-body simulation
  codes, but they suffer from a lack of versatility,
  i.e. a basis set has to be constructed for each
  different class of stellar system.
  We present  numerical techniques for generating a
  biorthonormal potential-density pair basis set
  which has a general specified pair as its lowest
  order member.  We go on to demonstrate how the set can
  be used to construct $N$-body equilibria, which we
  then evolve using an $N$-body code which calculates
  forces using the basis set.
\end{abstract}
%
%
%
\section{Introduction}
  $N$-body simulations of collisionless stellar systems generally
  require the evaluation
  of a smooth gravitational potential from a particle distribution
  containing fewer particles than stars in the real system
  by many orders of magnitude.
  One method for doing this originally proposed by
  Clutton-Brock (1972,1973) 
  involves the use of a set of basis functions which
  constitute potential-density pairs. The method has subsequently been used for
  stability studies of galactic equilibria 
  (e.g. Allen, Palmer \& Papaloizou 1990,
  Hernquist \& Ostriker 1992, Earn \& Sellwood 1995).
  The method expands the potential in terms of the specified basis set.
  The expansion coefficients are evaluated as integrals
  from the particle distribution
  which can be regarded as providing a Monte-Carlo estimate for them.
  The method relies upon the ability to accurately
  represent the initial equilibrium 
  distribution and its 
  subsequent evolution with the major contribution to the potential
  coming from the first few members of the basis set. These vary on the
  global length scale of the entire system and therefore
  have their expansion coefficients determined most accurately.

  Provided the specified basis set retains these advantages during
  the evolution of the system, the particles used to
  sample the distribution will evolve in a smooth potential, with the added
  computational advantage
  that the $N$-body problem is reduced to $N$ 1-body problems
  together with the evaluation of a relatively small number
  of expansion coefficients.  

  Earn and Sellwood (1995) performed a stability
  study of a thin disc in which
  they made a comparison of  normal mode growth 
  rates obtained using simulation techniques with
  values obtained from a linear perturbation analysis
  of Kalnajs (1976). They concluded  that the basis function method is the 
  optimal one for the study of the linear growth of
  instabilities in collisionless equilibria and therefore we suppose
  small departures from the equilibrium in general. However, they make the
  comment that the method
  is very specific. In particular the basis set has to be tailored
  to the particular density
  distribution under investigation, which is usually a difficult task and this
  may create problems when the density distribution changes significantly
  in a simulation.

  Up to now $N$-body simulations have been performed using
  basis sets which can
  be specified analytically and which are based on spherical or thin disc
  systems, though basis sets for more general axisymmetric systems
  have been developed (see Robijn \& Earn 1996, Earn 1996). 
  In this paper we investigate the possibility of numerically
  constructing a basis set which can in principle be tailored to
  any density distribution, although here we focus on axisymmetric systems.
  The idea is to construct the natural biorthonormal basis for the
  distribution
  such that the lowest order member corresponds to a specified system.
  The basis set is constructed by finding the eigenfunctions of a Fredholm
  integral equation numerically. These are used to provide
  the potentials, densities and forces of the basis set on a grid.
  As an illustration of the
  method we have determined the basis sets appropriate to a class
  of perfect oblate spheroids and Kuzmin-Toomre discs.
  The basis set generation code requires a  few hours
  CPU time to generate 40 basis
  potential-density pairs and the associated force functions on a dedicated
  workstation.  

  As we are interested in general distributions, the equilibria cannot
  in general be populated with particles by sampling a distribution
  function as this will be unknown. Instead we use an 
  orbit based method (see Schwarzschild 1979). This populates a
  time-average distribution of orbits which samples the mass distribution
  using a quadratic optimisation technique. 
  The subsequent departures from the specified equilibrium and
  relaxation of the $N$-body system obtained
  with evolution under the action of forces computed using the basis
  set are studied. In general the equilibrium virial ratio is maintained to
  within two percent using 20K particles  and 30 basis functions.
  Hernquist and Ostriker (1992) noted that discrete particle
  noise causes particle energies to undergo a random walk and thus
  leads to a relaxation of the system
  away from its original configuration. This effect is also studied
  and found to lead to a relaxation time of roughly order $N$ crossing
  times. Thus we have found that the basis sets we have
  constructed can be used effectively in simulations of the axisymmetric
  equilibria we have considered. They provide a promising tool for problems
  of stability  and other problems where the deviation
  from the initial configuration is not too large such as
  finding the response to weak external forcing
  and we shall report on these in the near future.

  In section 2 of this paper we describe the basic theory
  behind the construction of a customised basis set.
  In  section 3 we present details of the numerical construction
  of basis sets for two contrasting distributions by two different
  methods.
  In section 4 we go on to show how the basis sets can
  be employed to construct $N$-body representations
  of equilibrium stellar systems. In section 5 we
  evolve the equilibria using an $N$-body simulation code
  which uses the appropriate basis set for gravitational force calculation
  and perform various tests.  Finally in section 6 we discuss our results,
  and outline future work.
%
%
\section{Generating the basis set}
\subsection{Theory}
  Any potential-density pair $(\Phi_0,\rho_0)$
  satisfies Poisson's equation,
\begin{equation}
    \nabla^2 \Phi_0 = 4 \pi \rho_0,
\end{equation}
  where we have adopted units such that the 
  gravitational constant $G=1$.
  The solution is given by the Poisson integral
\begin{equation}
    \Phi_0 ( { \bf r }) = - \int \frac{\rho_0 ( {\bf r'}) }
    {\left| {\bf r-r'} \right|}d{V'}. \label{eq:poisson}
\end{equation}
  This and subsequent similar integrals are taken
  over the volume occupied by the mass distribution.

\noindent  Equation (\ref{eq:poisson}) can be rewritten as:
\begin{equation}
    \Phi_0 ( {\bf r} ) = \int  K ( {\bf r,r'})
       w ({ \bf r'} ) 
       \Phi_0 ( {\bf r'} ) dV',
      \label{eq:poisson2}
\end{equation}
  where the weight $w({\bf r})$ is given by
\begin{equation}
    w( {\bf r }) = - \frac {\rho_0 ( {\bf r })}
         { \Phi_0 ( {\bf r} ) } ,
\end{equation}
  and the kernel $K({\bf r,r'})$ by
\begin{equation}
    K({\bf r,r' }) = \frac {1}{ \left| \bf {r} - \bf {r'} \right| }.
\end{equation}
  Note that for an equilibrium $(\Phi_{0},\rho_{0})$
  pair, $w( {\bf r})$ is positive definite which is guaranteed for
  a density that is positive everywhere.  This
  is necessary for the method  given below to work.
  From (\ref{eq:poisson2}) it follows that
\begin{equation}
    \xi_0 ( {\bf r} ) = \int \overline{K} ( {\bf r,r'} )
     \xi_0 ( {\bf r'} )  dV',
     \label{eq:xi0}
\end{equation}
  where
\begin{equation}
    \xi_0 ( {\bf r} ) = \sqrt { w( {\bf r} ) } 
    \Phi_0 ( {\bf r } ),
      \label{eq:xidef}
\end{equation}
  and the symmetric kernel
\begin{equation}
    \overline{K}( {\bf r,r'} ) = \sqrt { w( { \bf r} ) w ({\bf r'} )}
    K( {\bf r,r'} ).
\end{equation}
  This means that $\xi_0$ is an eigenfunction of the 
  eigenvalue problem defined by the Fredholm integral
  equation (for a detailed explanation of the properties
  of this type of eigenvalue problem that we use below  see 
  Courant \& Hilbert, 1955, ch. 3),
\begin{equation}
    \lambda_k \xi_k ( {\bf r } ) = \int 
         \overline{K} ( { \bf r,r' } )
         \xi_k ({\bf r' }) 
         dV', \label {eq:xieq}
\end{equation}
  with eigenvalue $\lambda_0=1$.
  Given the eigenfunctions $ \{ \xi_k \} $
  we can calculate the $\{ \Phi_k \} $ via 
  (\ref{eq:xidef}), and the corresponding $ \{ \rho_k \}$
  are given by
\begin{equation}
    \lambda_k \rho_k( {\bf r } ) = -\Phi_k ( {\bf r } ) 
     w( {\bf r } ).
\end{equation}
  The variational properties of the eigenvalue problem,
  taken together with the fact that
  $ \overline{K} ( { \bf r,r' } )$ is always positive,  guarantee that
  the  lowest order
  eigenfunction which is associated with the largest eigenvalue 
   does not change sign and is therefore $\xi_0.$ The largest eigenvalue is  
  $\lambda_0=1$ and $\lambda_k \rightarrow 0$ for 
  $k \rightarrow \infty$.
  We define an inner product
\begin{equation}
    \langle  \Phi_k, \rho_l \rangle = - \int \Phi_k ( { \bf r } )
       \rho_l ( {\bf r } )   dV.
\end{equation}
  Then, when suitably normalised, the eigenfunctions
  obey a biorthonormality relation
\begin{equation}
    \langle \Phi_k, \rho_l \rangle = \delta_{kl}.
     \label {eq:biorth}
\end{equation}
  The completeness  properties of eigenfunction
  expansions of the type we consider described
  in Courant \& Hilbert (1955) chapters 3 and 5, indicate that
  the set is able to represent uniformly 
  and absolutely any potential $\Phi$ derived
  from the Poisson integral applied to a continuous
  density distribution that vanishes sufficiently rapidly
  at large  distances. Thus

\begin{equation}
    \Phi ( { \bf r } ) = \sum_{k} c_k \Phi_k ( { \bf r } ),
    \nonumber \end{equation}
 where
    \begin{equation}
    c_k = \langle \Phi_k , \rho \rangle =
    \langle \Phi, \rho_k \rangle. \end{equation}
  which is all that is needed for application to $N$-body calculations.
  Also we have the formal expansion for the density
\begin{equation}
    \rho ( { \bf r } ) = \sum_{k} c_k \rho_k ( { \bf r } ),
\end{equation}
  Thus given a density $\rho$ we can obtain
  the corresponding $\Phi$, and vice versa.
%
%
%
\subsection{Numerical solution of the eigenvalue problem}
  Generally equation (\ref{eq:xieq}) will not have an
  analytic solution and therefore numerical methods are
  required.
  We work in cylindrical polar coordinates $(R,z,\phi)$.
  Although in principle there are no symmetry
  restrictions on the equilibrium pair  $(\Phi_{0},\rho_{0}),$
  we shall in this paper assume them to be independent of $\phi.$
  Then the equilibrium is axisymmetric and the basis set can be separated
  into subsets with harmonic azimuthal dependence  associated 
  with a particular azimuthal mode number $m.$ 
  Thus we seek a
  representation of a given potential-density pair as follows;
\begin{eqnarray}
    \rho ( R,z, \phi )= \sum_{l,m}  \rho_{lm} ( R,z )
     \{ c_{lm} \cos m \phi + d_{lm} \sin m \phi \},
    \nonumber \\
    \Phi (R,z, \phi   ) = \sum_{l,m}  \Phi_{lm} ( R,z )
     \{ c_{lm} \cos m \phi + d_{lm} \sin m \phi \}.
    \label{eq:exp}
\end{eqnarray}

  \noindent That is, for a given $m$ we require basis pairs
  $(\Phi_{lm},\rho_{lm})e^{im \phi}$
  $l=0,1,2,\ldots$ which satisfy (\ref{eq:xieq}), with a different
  weight function $w_{m}=-\rho_{0m} / \Phi_{0m}$ for each value of $m$.

  We wish the lowest order pair $(\Phi_{00},\rho_{00})$
  to correspond to  the axisymmetric equilibrium.
  Thus we take $ \rho_{00} \equiv \rho_{0} $ to be the equilibrium
  density.

  To derive suitable $ \{\rho_{0m}:m\geq1\} $
  we begin by noting that  Poisson's equation for
  the axisymmetric equilibrium distribution is
\begin{equation}
    \xpp{\Phi_{00}}{R} + \frac{1}{R} \xp{\Phi_{00}}{R} +
      \xpp{\Phi_{00}}{z} = 4 \pi \rho_{00}.
\end{equation}
  Differentiating with respect to $R$ gives
\begin{eqnarray}
    \xpp{}{R} \left( \xp{\Phi_{00}}{R} \right) +
      \frac{1}{R} \xp{}{R} \left( \xp{\Phi_{00}}{R} \right) +
      \xpp{}{z} \left( \xp{\Phi_{00}}{R} \right) \nonumber \\
        -\frac{1}{R^2} \xp{\Phi_{00}}{R} =
        4 \pi \xp{\rho_{00}}{R},
\end{eqnarray}
  but this implies
\begin{eqnarray}
    \left( \xpp{}{R} + \frac{1}{R} \xp{}{R} +
      \xpp{}{z} + \frac{1}{R^2} \xpp{}{\phi} \right)
      \left( \xp{\Phi_{00}}{R} e^{i \phi} \right) = \nonumber \\
         4 \pi \xp{\rho_{00}}{R} e^{i \phi}.
\end{eqnarray}
  Hence $ \left(\xp{\Phi_{00}}{R},
     \xp{\rho_{00}}{R}\right)e^{i \phi}$
  is a potential-density pair with the $\phi$
  dependence required of a dipole $(m=1)$ term in (\ref{eq:exp}).
  It corresponds to a small translation of the system.
  Appropriate higher order $(\Phi_{0m},\rho_{0m})$ can be obtained
  after further differentiation with respect to $R$.
  One finds that a potential-density pair generated from 
  $ \left(\Phi_{00},\rho_{00}\right)$ is
  $ \left(R^m\left( {\partial^m\over \partial y^m}\Phi_{00}\right),
  R^m\left( {\partial^m\over \partial y^m}\rho_{00}\right)\right)
  e^{im\phi},$ where $y=R^2.$
  Functions derived in this way can be used to specify a pair
  $ \left(\Phi_{0m},
  \rho_{0m}\right)$ provided these are both of uniform sign
  as is the case for the examples considered below.
  But note that the procedure can also be applied to every member
  of a basis set for $m=0$ to generate a basis set for
  any non-zero $m.$  

  We present two alternative methods for solving the eigenvalue
  problem given by equation (\ref{eq:xieq}).  The first, method a),
  involves discretising the eigenvalue problem onto a grid and calculating
  the integral of (\ref{eq:xieq}) via direct summation, whilst the
  second, method b) uses a basis set built from standard orthogonal
  polynomials to represent the eigenvalue problem, the integration being
  calculated via Gaussian quadrature.
  Method a) has the advantage of in general being more accurate
  for a given amount of computer time, but method b) has more
  flexibility.
\subsection{Basis set generation via direct summation}
  We seek to numerically evaluate (\ref{eq:poisson})
  using the $ \{ \rho_{0m} e^{im\phi} \} $ as the density source terms.
  Because of the simple $ \phi $ dependence
  the $\phi$ integration can be expressed as a function of
  elliptic integrals, leaving
\begin{displaymath}
    \Phi_{0m}(R,z) = - \int_{R'=0}^{R'=\infty}
                       \int_{z'=-\infty}^{z'=\infty}
      \frac{F_m(k^2) }
      {\sqrt{(R+R')^2+(z-z')^2}}
\end{displaymath}
\begin{equation}
      \ \ \ \ \ \ \ \ \ \ \ \ \ \ \ \ \ \ \ \
      \ \ \ \ \ \ \ \ \ \ \ \  \times \rho_{0m}(R',z') R'dR'dz',
\end{equation}
  where
\begin{equation}
    k^2 = \frac{4RR'}{(R+R')^2+(z-z')^2},
\end{equation}
  and
\begin{equation}
    F_m (k^2) = 4 \int_{0}^{\pi/2} 
       \frac{e^{2im\phi}d\phi}{(1-k^2 \cos^2 \phi)^{1/2}}.
\end{equation}
  The $(R',z')$ integration
  is done by finite summation approximation on an
  $n~\times~n$ $(R_i,z_j)$ grid, i.e.
\begin{equation}
    [\Phi_{0m}]_{ij} = K_{ijpq} [\Sigma_{0m}]_{pq},
    \label{eq:finint}
\end{equation}
  where the summation convention is assumed, and
\begin{eqnarray}
    [\Phi_{0m}]_{ij} &=& \Phi_{0m} (R_i,z_j),
    \nonumber \\
    K_{ijpq} &=& \frac{-F_m(k^2_{ijpq})} 
      {\sqrt{(R_i+R_p)^2+(z_j-z_q)^2}},
    \nonumber \\
    k^2_{ijpq} &=& \frac{4 R_i R_p}
                        {(R_i + R_p)^2 + (z_j - z_q)^2},
    \nonumber \\
    \left[ \Sigma_{0m} \right] _{pq} &=& \rho_{0m} (R_p,z_q) R_p dR_p dz_q,
    \nonumber \\
    dR_p&=&\frac{1}{2}(R_{p+1}-R_{p-1}),
    \nonumber \\
    dz_q&=&\frac{1}{2}(z_{q+1}-z_{q-1}).
\end{eqnarray}
  The contribution to the integral of a cell in which
  ${\bf r = r'}$ is approximated by setting the value of
  the kernel in the cell to be the average value of the kernel
  in the four nearest neighbouring cells.

  In practice, especially when equilibria are not bounded
  in space, it is necessary to transform
  coordinates from the $(R,z)$ to the $(u,v)$ plane,
  say, where the exact choice of transformation is
  chosen to achieve maximum accuracy.  This leads to an
  extra Jacobian factor in $\left[ \Sigma_{0m} \right] _{pq}$,
  but does not affect the overall technique.
  
  Once this has been done the matrices are manipulated in the
  same manner as the original functions were in the
  theoretical case, giving the discrete analogue
  of (\ref{eq:xi0}),
\begin{equation}
    \lambda_{0m} [\xi_{0m}]_{ij} = \overline{K}_{ijpq}
      [\xi_{0m}]_{pq}.
     \label{eq:diseig}
\end{equation}
  By mapping the $n \times n$ grid to a vector with
  $n^2$ elements, equation (\ref{eq:diseig}) becomes a
  symmetric matrix eigenvalue problem which can be
  be solved, for example, using a $QL$ factorisation
  algorithm.  Thus the values of the first $n \times n$
  eigenfunctions on the coordinate grid are obtained,
  which can then be used for interpolation.

  The CPU time required to generate 40 basis functions and
  the associated forces is a few hours on a Sun Sparc20 
  workstation, which is small in comparison to the time required
  to conduct large long-term simulations.  This suggests that it
  may be possible to generate a new basis set midway through the
  simulation, if the system under study has evolved away from its
  initial state to such an extent that the old basis set is no
  longer able to accurately represent it.  
\subsection{Basis set generation via 
  Legendre polynomials and Gaussian quadrature} 
  The eigenvalue problem defined by (\ref{eq:xieq})
  can be expressed in matrix form using a polynomial basis.
  To do this we suppose a coordinate transformation
  that maps the cylindrical coordinates $(R,z)$ to $(u,v).$
  This is such that the infinite domain $0 < R <\infty,
  -\infty <z < \infty$
  is mapped into the finite domain $u_1<u<u_2, v_1<v<v_2.$
  For examples see below.

\noindent We now introduce the
  orthonormal  polynomials 
\begin{equation}
  {\cal P}_{ll'}\left(\mu_u,\mu_v\right)= N_{ll'}^{-1/2}
  P_{l}\left(\mu_u\right)
  P_{l'}\left(\mu_v\right),
\end{equation}
  for $l=0,1,2,...,l'=0,1,2,...,$ where $P$ 
  denotes the Legendre polynomial, with both
  $\mu_u = {(2u-u_1-u_2)\over
  (u_2-u_1)}$ and
  $\mu_v= {(2v-v_1-v_2)\over(v_2-v_1)}$ both lying in $(-1,1).$
  Here the normalization constant
  $N_{ll'}=(v_2-v_1)(u_2-u_1)/((2l+1)(2l'+1)).$
  These polynomials form an orthonormal basis over the $(u,v)$ domain.

  To obtain a matrix eigenvalue problem we first expand (e.g. for $m=0$)  the
  eigenfunction  $\xi_{k0}$ in the form

\begin{equation}
  \xi_{k0} =\sum_{ll'} {a_{k,ll'}{\cal P}_{ll'}\over \sqrt{|J|}},
  \label {pexp}
\end{equation}
  where $|J|$ gives the Jacobian such that
\begin{equation}
  RdRdz\rightarrow |J|dudv.
\end{equation}

\noindent Then the inner product
\begin{equation}
  \langle  \xi_{k0}, \xi_{k'0} \rangle = -2\pi
  \sum_{ll'} a_{k,ll'} a_{k',ll'} =-2\pi {\bf a}_k\cdot{\bf a}_{k'},
\end{equation}
  may be written as a scalar product of the column vectors 
  ${\bf a}_k$ and ${\bf a}_{k'}$  whose components are
  the expansion coefficients of $\xi_{k0}$ and $\xi_{k'0}.$

  Substituting the expansion (\ref{pexp}) into the eigenvalue problem
  defined through (\ref{eq:xieq}) and using the orthonormality of the
  polynomial basis expansion, we obtain the exactly equivalent matrix
  eigenvalue problem

\begin{equation}
  \lambda_k{\bf a}_k= {\bf M}\cdot{\bf a}_k,
\end{equation}
  where the symmetric matrix ${\bf M}$ has elements given by

\begin{eqnarray}
  {\bf M}_{ij}=\int {\overline{K} ({ \bf r,r' })}
  {\cal P}_i(u,v){\cal P}_j(u',v')\sqrt{|J|}\sqrt{|J'|} \nonumber \\
  \times d\phi dudvdu'dv'. 
  \label{melem}
\end{eqnarray}  
  Here we have adopted a mapping of the pairs of subscripts associated
  with the basis functions into single integers 
  $i$ and $j$ and $|J'|=|J(u',v')|.$

  Up until now the matrix representation of the
  eigenvalue problem is exact. Approximations to it are obtained
  by truncating the polynomial basis such that it becomes finite. 
  Thus we adopt $l=0,1,2...,L_{{\rm max}}-1,l'=0,1,2...,L_{{\rm max}}-1$
  giving $L_{max}$ functions in each of the coordinate directions
  or $L_{{\rm max}}^2$ in total.
  Solution of the now finite matrix eigenvalue problem,
  which can be accomplished with the QL algorithm as above,
  requires
  evaluation of the matrix elements given by $(\ref{melem}).$
  To do this requires evaluation of multiple integrals. We performed
  these using Gaussian quadrature in each of the $(u,v)$
  coordinate directions using $L_{{\rm max}}$ weights.
  This requires evaluation of the polynomial basis set at
  the coordinate points corresponding to the zeros
  of  $P_{L_{{\rm max}}}(\mu_u)$ and  $P_{L_{{\rm max}}}(\mu_v).$
  These points also provide a natural grid on which to evaluate
  the eigenfunctions once their expansion coefficients have been
  found from the solution of the eigenvalue problem.
  For practical evaluation, the kernel ${\overline{K} ({ \bf r,r' })}$
  was of necessity softened for ${\bf r}={\bf r}'$ by replacing $|z-z'|$
  by $0.05R$ rather than zero. Tests showed results to be independent
  of the details of the softening.

  We found that a successful solution to the eigenvalue problem
  and construction of the eigenfunctions for the examples given below
  could be obtained along these lines using black box software
  found in Press et al.  (1996).  Comparable results to those
  obtained with the grid based method, for  similar computational overhead,
  for the low order functions and performance in  $N$-body
  codes was obtained for $L_{{\rm max}}=30.$ Forces
  could be obtained
  as described below or by simple low order interpolation and
  numerical differentiation of the potentials.
%
%
%
  To conduct simulations using a basis set it
  is obviously necessary to possess the forces
  associated with the basis potentials.  If the
  potentials were known analytically, the simplest
  way to calculate the forces would be via direct
  differentiation.  But because the potentials are known only on a coordinate
  grid or via polynomial approximation it is more accurate to obtain the forces
  via integration of the densities over the force kernel, i.e.
  for the density function $\rho_{lm}(R,z)e^{im\phi},$
\begin{equation}
    { {\bf F}_{lm} ( { \bf r} ) } =
 e^{im\phi}\int \frac { ( { \bf r' - r } )e^{im(\phi'-\phi)} }
    { | { \bf r' - r } |^3 } \rho_{lm} ( R' ,z' ) dV',
\end{equation}
  using the same numerical integration algorithm that was
  used to calculate the potentials.
  Hence we obtain the set of forces $\{  {\bf F}_{lm} \}$ which
  correspond to the potential-density
  pair basis set $ \{ \Phi_{lm},\rho_{lm} \}$.

%
%
\section{Two example basis sets}
  As illustrations of the technique basis sets have been
  constructed for two contrasting distributions,
  the perfect oblate spheroid (POS), and
  a Kuzmin-Toomre (KT) disc with a Gaussian vertical
  density profile.  Henceforth all results shown were achieved
  using basis sets generated by method a), though similar results
  were obtained using functions generated via method b).
\subsection{Perfect oblate spheroid}
  The perfect oblate spheroidal potential (POSP) and 
  its corresponding density are best described 
  in some form of prolate spheroidal coordinates.  
  Details and properties of this distribution
  can be found in de Zeeuw (1985), and an analytic but non orthonormal
  set of functions was constructed for application to it by Syer (1995).
  The choice of coordinates made here for numerical integration 
  purposes was
\begin{figure*}
\centerline{\epsfig{file=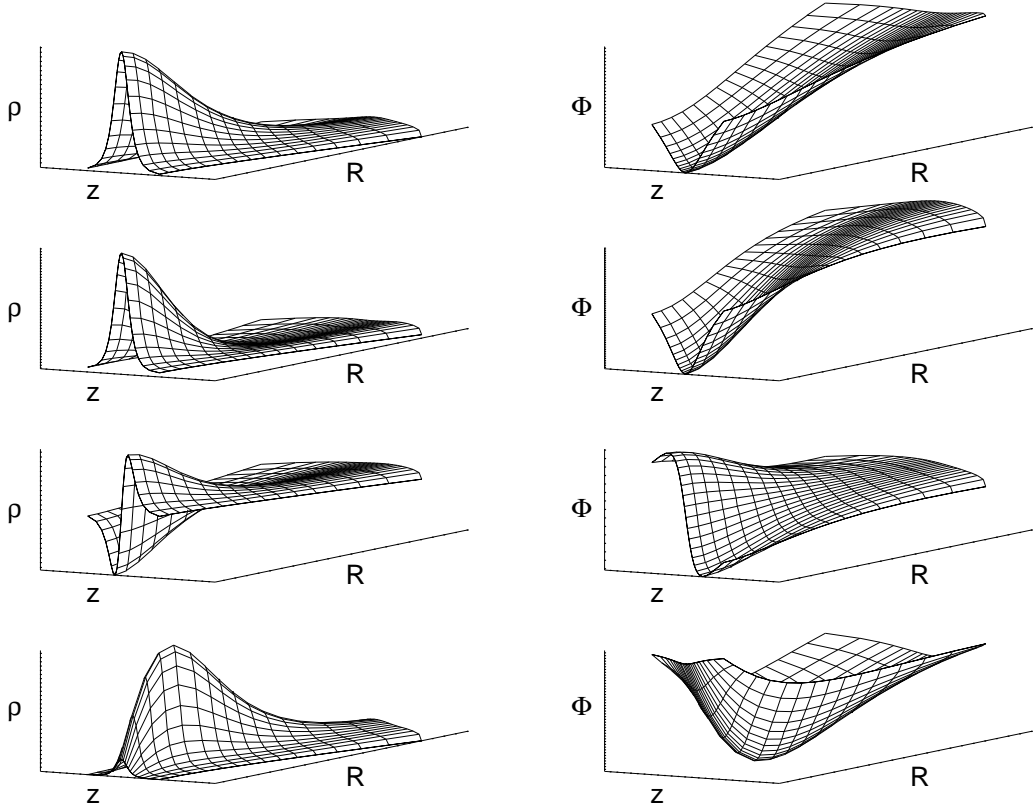,width=150mm}}
\caption{Some example $(\Phi_{lm},\rho_{lm})$ of the perfect 
  oblate spheroid with $e=5$.  From top to bottom they are
  $(l,m)=(0,0), (1,0), (5,0), (0,1).$}
\label{fig1}
\end{figure*}
\begin{eqnarray}
    z &=& \frac{1}{e} \tan u \tan v, \nonumber \\
    R^2 &=& \frac{1}{e^2} ( \tan^2 u - e^2 )(e^2 - \tan^2 v), 
      \nonumber \\
    u & \in &  [ \arctan (e), \pi/2 ), \nonumber \\
    v & \in &  [ -\arctan (e), \arctan (e) ],
\end{eqnarray}
  where $(u,v)$ are orthogonal coordinates in the $(R,z)$ plane.
  The parameter $e$ describes the flatness of the distribution
  (e.g. $e=0$ gives a spherical distribution, and the system
  flattens with increasing $e$). 
  Lines of constant $u$ give ellipses, lines of constant $v$
  give hyperbolae orthogonal to the constant $u$ ellipses
  and we are only concerned with a finite domain in the
  $(u,v)$ plane.

  The perfect oblate spheroidal
  density and potential are given by:
\begin{equation}
    \rho (u,v) = \frac {e(1+e^2)}{2 \pi} \cos^4 u \cos^4 v
    \equiv \rho_{00},
\end{equation}
\begin{equation}
    \Phi (u,v) = \frac {e (u \tan u - v \tan v) }
     { \tan^2 u - \tan^2 v }
     \equiv \Phi_{00}.
\end{equation}
\begin{figure*}
\centerline{\epsfig{file=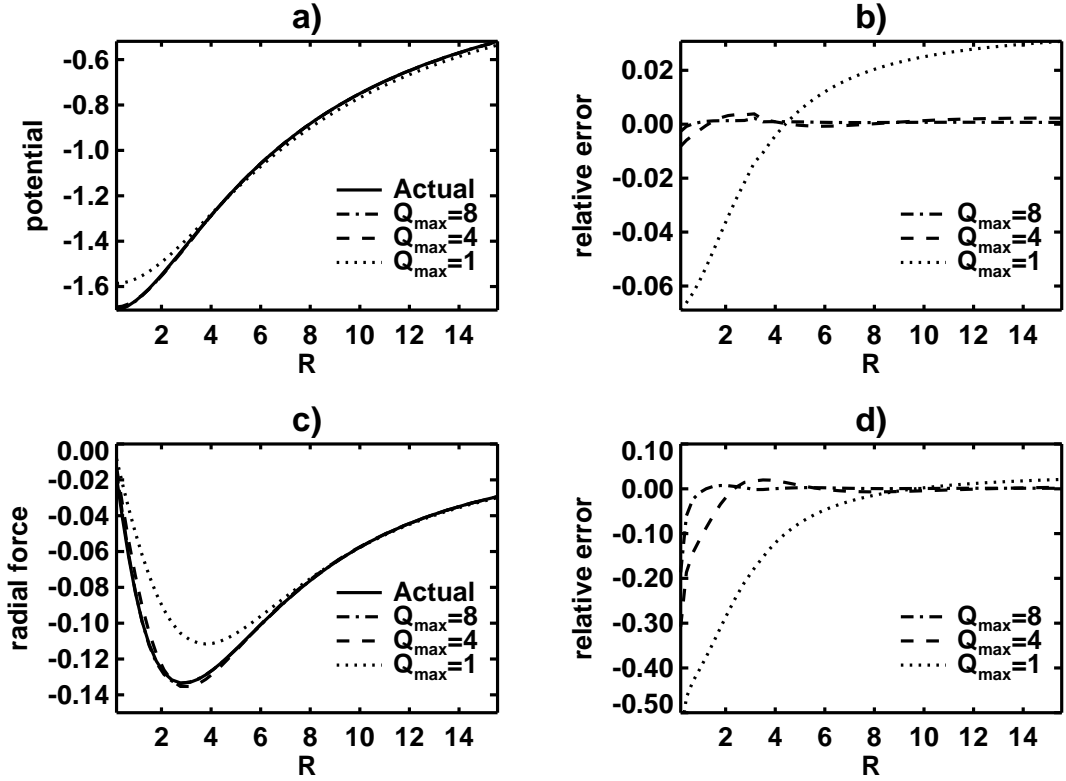,width=150mm}}
\caption{a) shows the basis representation of
  the potential given in equation~(\ref{eq:conv})
  in the $z=0$ plane using upto $Q_{\rm{max}}$ basis
  functions and b) shows the associated
  relative error, whilst
  c) and d) are the same plots, for the radial force
  expansion (for $z=0$).}
\label{fig2}
\end{figure*}
  Because the equilibrium potential is known 
  analytically, the numerically obtained $\{\Phi_{0m}\}$
  and their corresponding forces can be directly
  compared with the exact values and the accuracy of
  the numerical basis function calculation determined.
  We find that method a) using a $50\times50$ $(u,v)$ grid
  is sufficient to produce results with a maximum relative
  error in the potential of less than $0.2\%$ and in the
  force of $0.7\%$ over $90\%$ of the mass of the system,
  whilst method b) using $30\times30$ Gaussian weights has roughly
  double the error of the direct method.  This level of
  accuracy is more than sufficient
  for the purpose of $N$-body simulations where particle discreteness
  of the $N$-body system is the dominant source of error.

  Four of the potential-density pair basis functions for an $e=5$
  perfect oblate spheroid are shown in Figure \ref{fig1}.
  From top to bottom they are the original distribution,
  the first radially oscillatory function, the first
  $z$ oscillatory function, and the first dipole $(m=1)$ function.
\begin{figure*}
\centerline{\epsfig{file=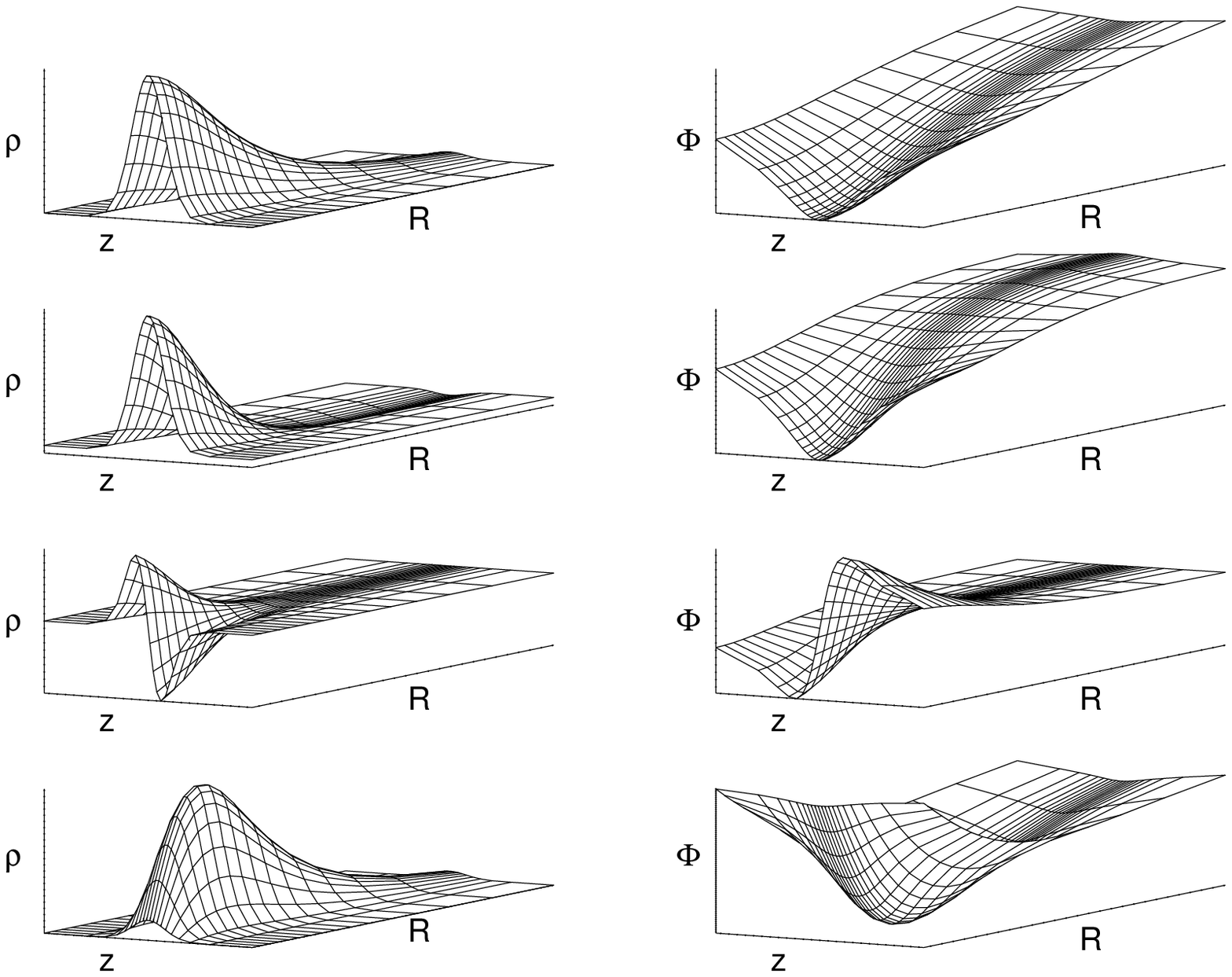,width=170mm}}
\caption{Some example $(\Phi_{lm},\rho_{lm})$ of the
  KT disc with $z_0=0.1, a=1, M=1$.
  From top to bottom they are $(l,m)=(0,0), (1,0), (5,0), (0,1)$.}
\label{fig3}
\end{figure*}
  Having constructed a basis set, we should
  ensure that the functions are capable of
  representing the kind of perturbations away from 
  equilibrium that we wish to investigate, using only
  the first few members of the set.
  We can illustrate this by representing
  the following potential (following Syer 1995),
\begin{equation}
    \Phi^{\rm pert}( {\bf r} ) = \Phi_1(  {\bf r}, e_1) +
      \alpha \Phi_2( {\bf r}, e_2),
      \label{eq:conv}
\end{equation}
  where $\Phi_1$ is the POSP that the basis set was
  based on, with eccentricity $e_1$, $\Phi_2$ is the POSP 
  of a different eccentricity $e_2$,
  and $\alpha$ is some positive number less than unity.
  (The integrations involved in evaluating the expansion
  coefficients were calculated numerically using the same
  coordinate grid on which the functions were originally
  generated).
  The convergence of the series is demonstrated graphically
  in Figure \ref{fig2}, for the values $ e_1=5, e_2=2, $ and 
  $ \alpha=0.3 $.  The left plots show successive 
  approximations of the potential and radial force 
  in the $z=0$ plane using 
  one, four, and eight basis functions and the right
  plots show the corresponding relative error (where we
  have defined the relative error of a quantity f, say,
  to be $f_{ \rm approx}/{f_{ \rm exact}} -1$).
  
  The high level of accuracy in the expansion representation
  of the perturbed distribution gives us confidence in
  the utility of our basis set.  (The error in the
  force expansion at small $R$ is not a problem in practice
  as the magnitude of the radial force tends to zero at
  small $R$ and it affects such a small fraction of the
  total mass of the system).
\subsection{Kuzmin-Toomre disc}
  The Kuzmin-Toomre disc profile with a gaussian
  vertical density profile (e.g. Sellwood 1994) is given by
\begin{equation}
    \rho(R,z) = \frac{\Sigma(R)}{(2\pi)^{1/2} z_0}
    \exp \left( {-\frac{z^2}{2z_0^2}} \right)
    \equiv \rho_{00},
    \label{eq:KTdens}
\end{equation}
  with
\begin{equation}
    \Sigma(R) = \frac{M}{2 \pi a^2}
      \left( {1 + \frac{R^2}{a^2} } \right) ^{-3/2}.
\end{equation}
  Again for numerical purposes we make a coordinate
  transformation in which the domain of interest is finite,
\begin{eqnarray}
    R & = & a \tan (u),
    \nonumber \\
    z & = & \sqrt{2} z_0 \tan(v),
    \nonumber \\
    u & \in & [ 0, \pi /2 ),
    \nonumber \\
    v & \in & ( -\pi / 2, \pi / 2 ).
    \label{eq:coord}
\end{eqnarray}
  The potential of the density given by (\ref{eq:KTdens}) 
  is not known analytically.  However, using a finite difference
  method on a  equally spaced $50 \times 50$ $(u,v)$ grid, the potential
  can be calculated to a specified accuracy. This can be compared with
  the potential obtained via the approximation (\ref{eq:finint})
  calculated on the same $(u,v)$ grid.  The maximum relative error
  is less than $1\%$ over 80\% of the mass of the disc using  method a),
  which is acceptable, particularly if the particle discs we use
  in simulations are truncated at this mass level.
  Similar accuracy was obtained with method b) with $L_{{\rm max}}=30.$
  Some members of
  the set with $z_0=0.1$, $a=1$, $M=1$ are shown in
  Figure \ref{fig3}.  From top to bottom they are the original distribution,
  the first radially oscillatory function, the first
  $z$ oscillatory function, and the first dipole $(m=1)$ function.
  We note that this system constitutes a severe test of the
  technique, because of the extreme differences in behaviour
  in the $R$ and $z$ directions.
%
%
%
\section{Construction of equilibrium models}
  The task is to assign positions and velocities to 
  $N$ discrete particles so that they represent a 
  steady state system described by the potential-density pair 
  $(\Phi_{00},\rho_{00})$.
\subsection{Orbit representation}
  The approach we follow is to write the density
  distribution of the particles as a weighted sum
  of contributions from $M$ different orbits, that is
\begin{equation}
    \rho ( { \bf r }, t ) = \sum_{k=1}^{M} m_k
      \delta ( { \bf r } - { \bf r }_k ( t )),
\end{equation}
  where each ${\bf r}_k (t)$ describes a different
  orbit labelled with the index $k.$ The total number of particles
  in each orbit is $m_k$ if each particle has unit mass (as is
  assumed throughout this paper).

  \subsection{Fitting the prescribed density}
  Using the first $Q_{\rm{max}}$ axisymmetric members of the
  basis set constructed earlier, for any density
  $\rho ( { \bf r }, t )$ we can construct the quantity   

\begin{equation}
    D(t) = \langle \Phi_{00} - \sum_{l=0}^{Q_{\rm{max}}-1} 
            c_{l0}(t) \Phi_{l0},
            \rho_{00} - \sum_{l=0}^{Q_{\rm{max}}-1} c_{l0}(t) 
            \rho_{l0} \rangle ,
\end{equation}
  where
\begin{equation}
    c_{l0} (t) = \langle \Phi_{l0}( {\bf r} ), 
               \rho( {\bf r},t ) \rangle
           = - \sum_{k=1}^{M} m_k \Phi_{l0} ( { \bf r }_k (t)).
\end{equation} 
  Using the biorthonormality relation (\ref{eq:biorth}),
  $D$ can be written as:
\begin{equation}
    D = 1 - 2 {c}_{00} + 
         \sum_{l=0}^{Q_{\rm{max}}-1} {c}_{l0}^2,
\end{equation}
  We see that $D$ is minimised when ${c}_{00}=1$, and 
  ${c}_{l0}=0 \ \ {\rm for } \ \  l \ge 1 $.

  Because $D$ is non-negative and is zero only when
  $ \rho ( { \bf r }, t )= \rho_{00}( {\bf r},t )$, we can regard it,
  in an appropriate norm, as
  a measure of the `distance' that $ \rho( {\bf r},t ) $
  is `away from' $ \rho_{00}( {\bf r}) .$

  Further, as  we are seeking a steady state solution
  which should have a time independent
  density, we can take a long-time average to
  remove the explicit $t$ dependence
  in the coefficients $c_{l0}(t).$
  Then we find

\begin{equation}
    D = \langle \Phi_{00} - \sum_{l=0}^{Q_{\rm{max}}-1} \overline{c}_{l0}
         \Phi_{l0}, \rho_{00} - \sum_{l=0}^{Q_{\rm{max}}-1} 
         \overline{c}_{l0} \rho_{l0} \rangle,
\end{equation}
  where each coefficient is replaced by its time average 
  denoted with an overline
  $$\overline{c}_{l0}=\lim_{t \rightarrow \infty}
    \left\{{\int^t_0 c_{l0}(t) dt \over t} \right\}.$$
  In practice we have found that stable time averages 
  can be obtained after following an
  orbit for a time of order a hundred
  crossing times for the cases we have considered.

  The task is to assign a different mass to
  each orbit to make the distribution of the particles match
  $\rho_{00}$ as closely as possible,
  in a time averaged sense.
  This can be regarded as being
  equivalent to minimising $D$ subject to the constraint
  that $m_k \ge 0.$
  Using the time averaged coefficients,
  $D$ can be written as:
\begin{equation}
    D = 1 - 2 \overline{c}_{00} + 
         \sum_{l=0}^{Q_{\rm{max}}-1} \overline{c}_{l0}^2,
\end{equation}
  with
\begin{equation}
    \overline{c}_{l0} = - \sum_{k=1}^{M} m_k 
          \overline{\Phi}_{l0}( {\bf r}_k ),
\end{equation}
  and
\begin{equation}
    \overline{c}_{l0}^2 = \sum_{k,p=1}^M m_k m_p 
           \overline{\Phi}_{l0}({\bf r}_k)
           \overline{\Phi}_{l0}({\bf r}_p).
\end{equation}

We may write this in the equivalent form
\begin{equation}
    D=1 + 2 {\bf m \cdot \overline{\Phi}}_{00}
      + {\bf m} \cdot {\bf A} \cdot {\bf m},
\end{equation}
  where the vector components $\left[{\bf m}\right]_k=m_k,$ and
  $\left[\overline{\bf {\Phi}}_{00}\right]_k=\overline{\Phi}_{00}({\bf r_k}),$
 with  the matrix element

\begin{equation}
    {\bf A}_{kp} = \sum_{l=0}^{Q_{\rm{max}}-1}
             \overline{\Phi}_{l0} ( {\bf r}_k )
             \overline{\Phi}_{l0} ({\bf r}_p),
              \hspace{0.25in} k,p=1, \ldots, M.
\end{equation}
  Thus, to calculate ${\bf m}$ in such a way as to provide
  the closest fit to the specified equilibrium, 
  we have a quadratic optimisation problem, namely to
  minimise $D$ subject to the constraint 
  $m_k \ge 0,  k=1,...,M.$
\newline

  However there is a further complication, as the number
  of eigenfunctions used is (usually considerably) smaller than
  the number of orbits, making the problem ill-conditioned, 
  and the particular method we use to minimise $D$ gives
  a solution where very few occupation numbers are
  non-zero.  This ill-conditioning
  can be removed by the use of a profit function.
  By adding a function to $D$
  which is small when the distribution of orbits is smooth, the
  degeneracy of the solution is removed, and a wider
  spread of occupation numbers is achieved (For examples, see
  Richstone \& Tremaine 1988, Merritt 1993,
  Syer \& Tremaine 1996).  We can interpret
  the solution to the modified minimisation problem as one which
  minimises the deviation from the desired density distribution
  whilst simultaneously using surplus degrees of freedom manifest
  in the degeneracy of the pure problem to minimise the profit function.

  We note that Tremaine,
  H\'{e}non \& Lynden-Bell (1986) argue that in the presence
  of small scale mixing, all functions of
  the form $-\int C(F) d{\bf x}d{\bf v},$ where $F$ is
  the distribution function and $C$ is any convex function,
  increase during collisionless relaxation, though in
  general this conjecture is false (Sridhar 1987).
  This suggests that we seek a profit function of the form
  $\sum_{k=1}^{M} C(m_k).$
  In this context, if phase space is divided up into
  cells of unit volume, and the above integral  approximated
  by a sum, the $\{m_k\}$ may be regarded as 
  the discrete equivalent of the distribution function.
  The quadratic nature of the original $D$ motivates us to
  construct a profit function which is likewise quadratic
  as the nature of the optimisation problem then remains
  the same. Hence we have taken the profit function to be of the form
\begin{equation}
    S = \sum_{k=1}^{M} \left( \frac{m_k}{p_k} -1 \right) ^2,
\end{equation}
  where the $\{ p_k \}$ are specified statistical weights.  In practice a
  reasonable choice for these is such as to give each orbit equal
  weight, e.g.
\begin{equation}
    p_k = M.
\end{equation}
  Proceeding as outlined above, we
  form a new function $D'$ to be minimised such that
\begin{equation}
    D' = D + \mu S,
\end{equation}
  where $\mu$ is a small positive constant whose magnitude reflects
  the degree of smoothness imposed on the solution.
  By solving this quadratic optimisation problem the
  required $\{ m_k \}$ for an equilibrium solution
  can be calculated. We use the two phase quadratic programming
  method of Gill et al.  (1986)  
  which is implemented in the NAG  library routine
  E04NCF. 
  On each orbit $k$ we then place $n$ 
  particles, where $n$ is proportional to $m_k.$  
  By placing the particles at equal time intervals
  along each orbit we ensure a smooth distribution in orbital
  phase.
\subsection{Application to the perfect oblate spheroid}
  The POSP, being a St\"{a}ckel potential, is completely
  integrable. Each orbit has three conserved actions
  associated with it
  making the construction of an $M$-orbit family on which
  to base our equilibrium easier. But the method can be
  applied to more general mass distributions, as will be seen
  in the next section, provided an $M$-orbit family
  can be constructed which adequately samples phase space. 

  As a consequence of integrability, a general
  orbit in the POSP separates into librations between
  fixed values of the $u$ and $v$ coordinates, so that
  orbits can be classified by the triple $(u_+,u_-,v)$
  where $u_+,u_-$ are the extrema in the $u$ coordinate
  and $\pm v$ the extrema in the $v$ coordinate.
  In terms of a thin disc model we can 
  think of oscillations in $u$ and $v$ as corresponding to radial
  and vertical oscillations respectively.
  Here we seek an equilibrium distribution which consists 
  of stars on infinitesimally
  thin short-axis tube orbits (a ``shell orbit'' model).  These
  orbits have no libration in the $u$ direction. Thus, in the thin
  disc limit  they correspond
  to circular orbits with superposed small amplitude vertical oscillations.
  Bishop (1987) tackled this problem via a numerical
  solution of the corresponding density/distribution
  function integral equation.

\begin{figure}
\centerline{\epsfig{file=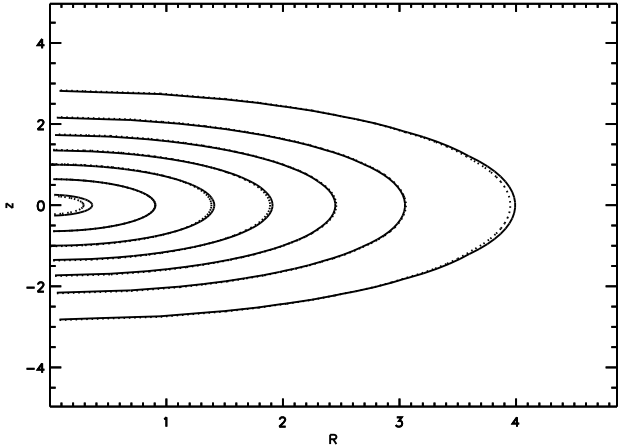,width=90mm}}
\caption{Comparison of the analytic density (solid line)
  with that of an $N$-body sample (dashed line) for
  an $e=1$ perfect oblate spheroid.}
\label{fig4}
\end{figure}
  To set up an orbit family we divide the $(u,v)$ plane
  into a grid which evenly samples the mass of the system
  and then allocate an orbit to each grid cell which
  has its turning points randomly allocated within the cell.
  These orbits are then integrated in the POSP
  for such a time that the $\overline{\Phi}_{l0}$ are 
  determined to a sufficient accuracy.

  Using an orbit family of $1000$ orbits together with $40$ of the 
  $ \Phi_{l0} $  and applying no smoothing
  (i.e. $\mu = 0$), a good $N$-body realisation 
  of an equilibrium $e=1$ perfect 
  oblate spheroid (in the sense of a
  small value of $D'$) is found but only around
  $100$ orbits are occupied.  However, setting $\mu=10^{-4}$
  increases the number of occupied orbits to around 800 whilst
  having little effect on the value of $D'$.
  A contour plot comparing the analytic density with the 
  basis set representation of the $N$-body distribution
  (using $100$K particles) is shown in Figure \ref{fig4}
  (where the contours are logarithmically spaced).
\subsection{Application to the KT disc}
  The orbits we use to build our KT disc from are taken
  by sampling one of the distribution functions used 
  in Sellwood \& Merritt (1994).
  They used an analytic distribution function for the planar
  motion (derived in Kalnajs 1976).  They assigned $z$ velocities by first
  integrating the one-dimensional Jeans equation for the vertical motion
\begin{equation}
  \frac{1}{\rho} \xp{ (\rho \sigma_{w}^2)}{z} =  F_{z}
  \label{eq:zjeans}
\end{equation}
  (where $F_z$ is the vertical component of the gravitational
  acceleration) 
  to find  the  velocity dispersion $\sigma_{w}(z)$, and then 
  by assuming the vertical
  velocity distribution to be a Gaussian of width $\sigma_{w}(z).$
  We generated our orbit
  family by sampling the distribution function in two stages.
  First we sampled values of $R$ and $z$ from the density
  distribution $\rho(R,z).$ Then for each particle position
  we sampled  the velocity component
  perpendicular to the vertical $z$ axis using the Kalnajs distribution
  function.  Finally the vertical velocity component
  $v_{z}$ was sampled from the Gaussian of width $\sigma_{w}(z),$
  the latter having been determined from 
  (\ref{eq:zjeans}).   Positions and velocities determined in the above
  manner were used to provide the initial conditions for the orbit family.

\begin{figure}
\centerline{\epsfig{file=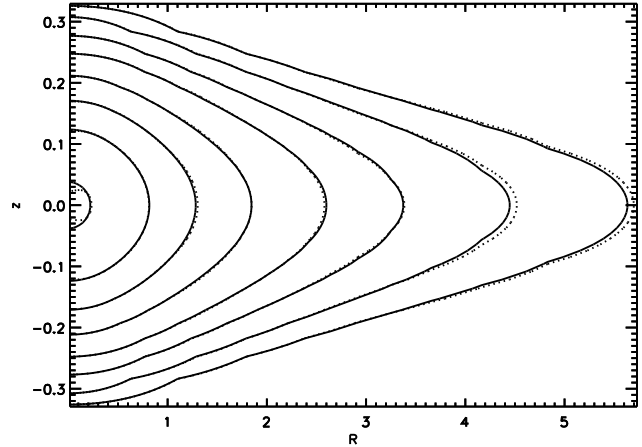,width=90mm}}
\caption{Comparison of the  analytic density (solid line)
  with that of an $N$-body sample (dashed line) for
  $z_0=0.1$ KT disc.}
\label{fig5}
\end{figure}

  We adopted the planar distribution function
  which is independent of angular momentum, and hence isotropic
  $(m_{K}=3$ in the notation of Sellwood \& Merritt 1994),
  and take the $z$ scale height $z_0=0.1$
  As before an orbit family of $1000$ orbits and $40$ of
  the $ \Phi_{l0} $ were  used, and
  setting $\mu = 10^{-3}$ resulted in around $900$ of these
  orbits being occupied.
  A contour plot comparing the analytic density with the 
  basis set representation of the $N$-body distribution
  (using $100$K particles)
  is shown in Figure \ref{fig5} (again with logarithmically
  spaced contours).
%
%
%
\section{Equilibrium simulations}
  Our $N$-body simulation code uses a standard second-order
  time-centred leapfrog integration scheme.  The forces
  on each particle are calculated from the basis set expansion
\begin{equation}
  {\bf F} ( {\bf r}_k ) =
       - \sum_{l,m} c_{lm} {\bf F}_{lm} ({\bf r}_k),
\end{equation}
  where the fourier coefficients $c_{lm}$ are calculated from
  the set of particle positions $\{{\bf r}_k\}$ via
\begin{equation}
  c_{lm} = -\sum_{k=1}^{N} \Phi_{lm} ({\bf r}_k),
\end{equation}

  To assess the quality of our simulations, we conduct energy
  and angular momentum conservation tests and calculate
  the relaxation rate for the POS and KT disc equilibria
  generated in the previous section. We use $20K$ particles,
  $30$ of the basis functions, and a step-size of approximately
  $0.004$ times the dynamical time
  $t_{\rm{dyn}}$ of each system, where we have defined
  $t_{\rm{dyn}}$ to be the time for a particle at the
  half-mass radius of the system to complete a quarter of
  one circular orbit around the system.
  We adopt the appropriate $t_{\rm{dyn}}$ as the
  unit of time in both cases.  The results presented here were
  produced using functions generated via method a), though method
  b) gives similar results.

  To suppress any instabilities we impose axisymmetry and
  reflection symmetry in the $z=0$ plane, by omitting all
  $m>0$, and z antisymmetric functions respectively
  from our basis set expansions.
  We also ran the systems
  for several  time units before beginning the tests
  to allow any initial fluctuations to phase mix away.
  This ensures that we are testing the
  quality of the code, rather than discreteness effects in the
  initial equilibria.  We give results for the simulations
  over a period of ten dynamical times, though we ran the
  systems for many more dynamical times  to ensure that the results
  given here are typical.

  Sellwood \& Merritt (1994) advocated the use of a quiet start
  in their $N$-body stability studies of KT discs.  One feature
  of their quiet start is that they allocated
  particles in groups obeying certain symmetries
  (e.g. in the $z$ direction and in azimuthal segments).
  This ensures that the net linear and angular
  momentum of the system is zero and that low azimuthal mode numbers
  can be made to be absent from the initial density distribution.
  It also ensures that any subsequent rapid
  growth of such an $m\ne 0$ component
  can be ascribed to instabilities.  However, since here
  we are testing axisymmetric equilibria simulations and effectively
  impose such symmetries on the system {\em ab initio} by excluding
  all $z$ antisymmetric and non-axisymmetric functions from the
  basis expansions such a quiet start becomes redundant (except
  in the total momentum conservation tests).
  In fact we find that in this situation
  using a quiet start may  raise the noise level and relaxation
  rate, since, for example, allocating particles in pairs, at
  phase positions $(x,y,\pm z,v_{x},v_{y},\pm v_{z})$ effectively
  halves the number of particles the expansion can `see' and
  thus raises the noise level by a factor of $\sqrt{2}$ since
  noise scales as $1/ \sqrt{N}.$
\subsection{Virial ratio}
  The scalar virial theorem (see, e.g. Binney \& Tremaine 1987)
  tells us that
\begin{equation}
  \frac{1}{2} \ddot{I} = 2 {KE}_{ \rm{total}} +
      {PE}_{ \rm{total}},
\end{equation}
  where
\begin{equation}
  I = \sum_{i=1}^{N} \left| { \bf x}_i \right|^2 ,
  \nonumber \\
\end{equation}
\begin{figure}
\centerline{\epsfig{file=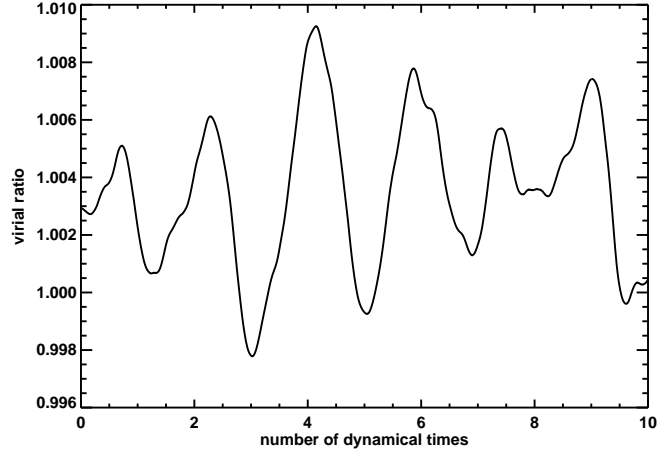,width=90mm}}
\caption{Evolution of virial ratio of POS equilibrium simulation.}
\label{fig6}
\end{figure}
\begin{figure}
\centerline{\epsfig{file=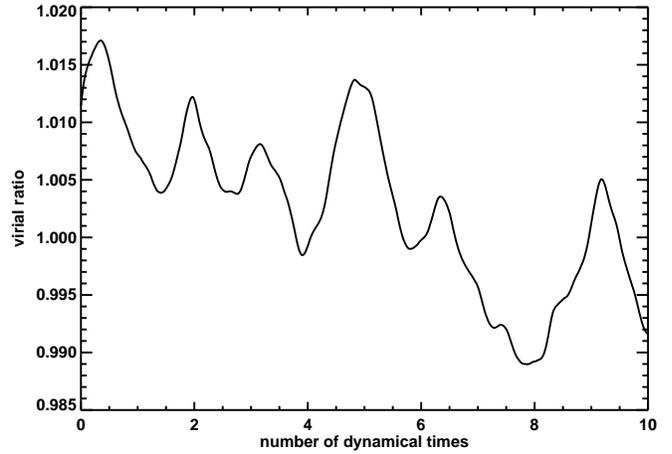,width=90mm}}
\caption{Evolution of virial ratio of KT disc equilibrium simulation.}
\label{fig7}
\end{figure}
\begin{eqnarray}
  {KE}_{ \rm{total}} &=& \frac{1}{2} \sum_{i=1}^{N}
      \left| {\bf v}_{i} \right|^2,
    \nonumber \\
  {PE}_{ \rm{total}} &=&-\frac{1}{2} \sum_{i,j=1,i\ne j}^{N}
     \frac{1}{\left| {\bf x}_{i} - {\bf x}_{j} \right| } 
      \nonumber \\
                    &=& - \frac{1}{2} \sum_{l,m} c_{lm}^2.
\end{eqnarray}

  In exact steady state equilibrium with an infinite number of basis
  functions, the quantity $\ddot{I}$
  should be identically zero.  Thus the behaviour of
  $-2 {KE}_{ \rm{total}} / {PE}_{ \rm{total}}$ is an indication
  of how close our $N$-body systems are to true equilibrium.
  As shown in Figures \ref{fig6} and \ref{fig7}
  the virial ratio in both systems is maintained at
  unity $\pm 0.02$ which is consistent with the
  statistical error we would expect with $20K$ particles.
\subsection{Total energy conservation}
  The total energy of the system is given by
\begin{equation}
  E_{ \rm{total}} = KE_{ \rm{total}} + PE_{ \rm{total}},
   \nonumber
\end{equation}
  The fluctuation in $E_{ \rm{total}}$ is less than 0.1 \%
  in our simulations.
\begin{figure}
\centerline{\epsfig{file=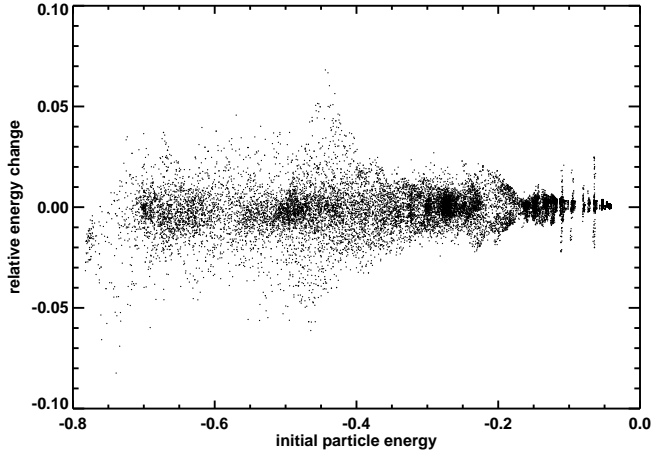,width=90mm}}
\caption{Individual particle energy deviations for the POS equilibrium simulation.}
\label{fig8}
\end{figure}
\begin{figure}
\centerline{\epsfig{file=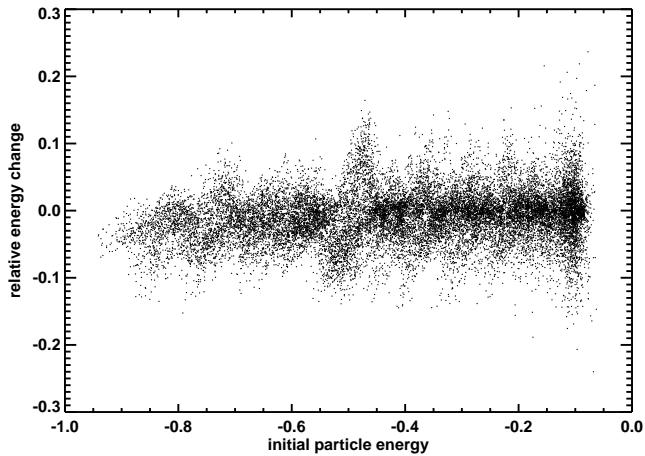,width=90mm}}
\caption{Individual particle energy deviations for the KT disc equilibrium simulation.}
\label{fig9}
\end{figure}
\subsection{Relaxation}
  In a collisionless system, individual particle
  energies should be conserved.  In our simulations,
  however, due to discreteness noise individual particle energies
  will tend to random walk slightly with time - i.e. the system
  relaxes away from the initial collisionless equilibrium.
  This effect is shown in Figures \ref{fig8} and \ref{fig9}
  where we plot the relative change in energy of the $N$
  particles over the duration of the test of $10 t_{\rm{dyn}}$
  against their initial energies.  The vertical band-like structure
  observed in these scatter plots is found in similar simulations
  described in Vine \& Sigurdsson (1998).  They state that there
  will be regions where the discrepancy between the value of the potential
  given by the basis expansion and the true value of the potential is greater
  than average (because of the  expansion truncation), and that
  the bands contain the particles which occupy these regions.

  To quantify the relaxation we define
\begin{equation}
  A = \frac{1}{N} \sum_{k=1}^{N} 
     \left[ \frac{E_k(t)-E_k(t_0)}{E_k(t_0)} \right]^2,
\end{equation} 
  where $E_{i}(t)$ is the energy of particle $i$ at time $t$.
  We note that if the energies of individual particles undergo
  a random walk, $A$ as well as each individual
  term in the summation is expected to increase linearly with time.

\begin{figure}
\centerline{\epsfig{file=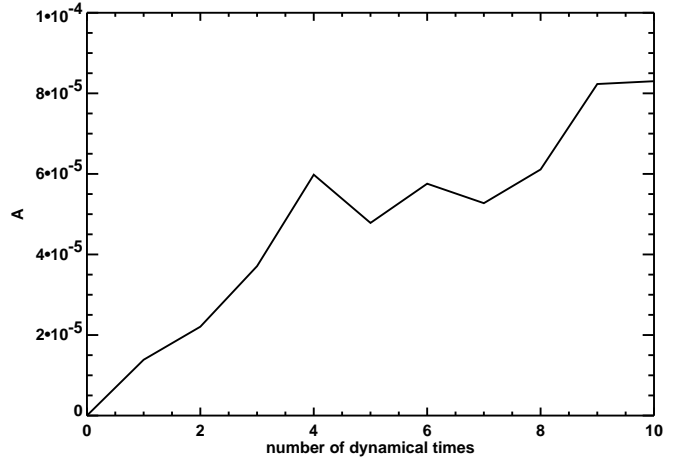,width=90mm}}
\caption{Relaxation of POS equilibrium simulation.}
\label{fig10}
\end{figure}
\begin{figure}
\centerline{\epsfig{file=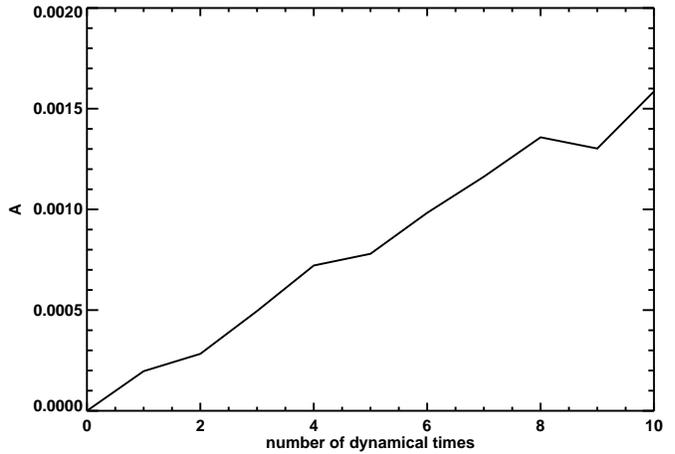,width=90mm}}
\caption{Relaxation of KT equilibrium simulation.}
\label{fig11}
\end{figure}
  In Figures \ref{fig10} and \ref{fig11} we
  can see that indeed $A$ grows linearly in
  time as expected.  However, the  rate of increase
  is different for the two systems.  There are at
  least two possible explanations for this.
  The first is that
  the definition of $t_{\rm{dyn}}$ is fairly arbitrary and
  is only an indicator of the dynamical time-scale of
  the two systems, hence we may not be exactly comparing
  like with like.
  The second, which may explain why the relaxation rate
  of the POS simulation is smaller than that of the KT disc
  is due to the different nature of the orbits in the two
  systems.  The KT disc model contains many low
  angular momentum, radial orbits which pass close to
  the centre of the system.  It is likely that such
  orbits will suffer more relaxation on average as
  the centre of the system is subject to proportionately
  more noise disruption.  With the POS ``shell orbit''
  model fewer particles pass near the centre of the
  system, and the orbits are more isolated from each
  other in phase space, so we might expect the relaxation
  rate in this model to be smaller.

  For the POS simulation the relaxation time is of
  order $10^5$ times $t_{\rm{dyn}}$,
  whilst for the KT disc it is of the order $10^4$
  times $t_{\rm{dyn}}$, hence we can run simulations
  for hundreds of dynamical times before becoming too concerned
  about the effects of relaxation.  The relaxation
  rate for runs using the basis set generated by method b) with
  $L_{{\rm max}}=30$
  has about a $10\%$ larger relaxation rate, because of the lower
  accuracy of the method b) functions.

  Runs with different numbers of particles indicate that
  the relaxation rate is inversely proportional to the number of
  particles used, as we would expect.

  We also briefly study the dependence of the relaxation
  rate on the number of basis functions used.
  A simulation using only the zeroth order basis function has
  a much reduced relaxation rate (by around a factor of $10$),
  but a simulation using the first five
  $z$ and axisymmetric functions has much the
  same relaxation rate as the original simulation which
  employs thirty.  This effect can be understood from Weinberg (1993),
  which predicts that the relaxation rate of an $N$-body simulation
  code in which small scale particle fluctuations have been
  suppressed will be dominated by large scale modes which are
  generated by particle noise and amplified by self-gravity.  In our
  results the runs with five functions adequately represent
  these dominant modes and additional functions contribute little
  to the rate of relaxation.

\subsection{Linear and angular momentum}
  Any simulation code in which the forces are not
  calculated via pair-wise summation will violate
  momentum conservation to some degree.  The drift
  in linear momentum can be compensated for by
  recentreing the coordinate grid on the centre of mass of the
  system.  By imposing suitable symmetries on the
  initial particle distribution its total angular
  momentum ${\bf L}= (L_x,L_y,L_z) $ is set to zero.  During the simulations
  $L_z$ is conserved exactly for individual particles
  (which is imposed by the exclusion of non-axisymmetric
  forces), hence total $L_z$ is also conserved exactly.
  Whilst individual $L_x$ and $L_y$ should not be conserved, the
  total $L_x$ and $L_y$ are seen to be conserved exactly
  in our simulations.
%
%
%
\section{Conclusions and further work}
  We have shown that is possible to numerically construct
  a potential-density pair basis set which is customised
  to a general equilibrium distribution, and demonstrated
  how the set can represent deviations from
  the original distribution with a high level of accuracy.

  We have demonstrated how the set can be used to construct
  an $N$-body equilibrium sample of a density distribution without
  explicit knowledge of the integrals of motion of the
  system and without explicit reference to a known distribution
  function using an orbit based method. We have developed
  an $N$-body simulation code which calculates particle
  forces using the basis set, and have demonstrated
  the accuracy of the code with regards to conservation
  of integrals of motion and relaxation effects, and described
  the effect of changing the number of particles and
  basis functions used.

  Because the basis construction is
  completely general, we can apply the technique to systems
  for which no suitable analytic basis set is known.
  Since the CPU time required to generate a new basis set is
  not prohibitive,  it should also prove
  possible to follow the evolution
  of a given system even if it has evolved significantly
  away from its original configuration by generating a
  new basis set from the current density distribution.

  Because the computational cost of the simulation
  code scales linearly with $N$, large numbers
  of particles can be employed which will reduce the effect
  of particle noise resulting in more accurate simulations.
  There has been much discussion on the relative merits of
  different $N$-body simulation methods (e.g.
  Hernquist \& Barnes 1990, Hernquist \& Ostriker 1992,
  Earn \& Sellwood 1995, Sellwood 1997) and it seems
  clear that for certain classes of problem,
  namely detailed studies  that involve not too large deviations from an
  initial distribution, a basis set expansion
  $N$-body code is one of the most competitive.

  We comment further that the method can be applied to problems
  involving two (or more) distinct mass distributions, for example
  a galactic disc embedded in a dark matter halo system. This involves
  combining the kind of axisymmetric basis set described
  in this paper to represent the disc distribution with a
  similarly generated spherical basis set to represent the
  halo distribution.  In addition to stability studies, problems
  involving the warping of a galactic disc embedded in a
  halo, or the effects of an infalling satellite on the system
  can be studied.  We hope to report on such work in the near future.
%
%
\section*{ACKNOWLEDGMENTS} This work was supported by PPARC grant GR/H/09454,
  and MJWB is supported by a PPARC studentship.
\section*{References}
  Allen, A.J., Palmer, P.L., \& Papaloizou, J. 1990, MNRAS,
  242, 576 \\
  Binney, J., \& Tremaine, S. 1987, 
    Galactic Dynamics (Princeton Univ. Press) \\
  Bishop, J., L. 1987, ApJ, 322, 618 \\
  Clutton-Brock, M. 1972, Ap\&SS, 16, 101 \\
  Clutton-Brock, M. 1973, Ap\&SS, 23, 55 \\
  Courant, R., \& Hilbert, D. 1955, Methods of
    Mathematical Physics Vol I(2nd ed.; New York:Interscience Publishers) \\
  de Zeeuw, P. T., 1985, MNRAS, 216, 273 \\
  Earn, D. J. D., 1996, ApJ, 465, 91 \\ 
  Earn, D. J. D., \& Sellwood, J. A. 1995, ApJ, 451, 533 \\
  Gill, P.E., Hammarling, S.J., Murray, W., Saunders, M. A. \& Wright, 
    M. H., 1986, User's guide for LSSOL (Version 1.0): A Fortran package for
    constrained linear least-squares and convex quadratic programming.
    Technical Report SOL 86-1. Systems Optimization Laboratory, Dept.
    of Operations Research, Stanford Univ. \\
  Hernquist, L., \& Barnes, J. E. 1990, ApJ, 349, 562 \\ 
  Hernquist, L., \& Ostriker, J. P. 1992, ApJ, 386, 375 \\
  Kalnajs, A. J., 1976, ApJ, 205, 751 \\
  Merritt, D., 1993, ApJ, 413, 79 \\
  Press W.H., Flannery B.P., Teukolsky S.A., \& Vetterling W.T., 1996,
    Numerical Recipes: The Art of Scientific Computing, Cambridge
    University Press\\
  Richstone, D. O., \& Tremaine, S. 1988, ApJ, 327, 82 \\
  Robijn, F. H. A., \& Earn, D. J. D. 1996, MNRAS, 282, 1129 \\ 
  Schwarzschild, M. 1979, ApJ, 232, 236 \\
  Sellwood, J. A. 1997, Computational Astrophysics
    (ASP Conference Series) \\
  Sellwood, J. A., \& Merritt, D. 1994, ApJ, 425, 530 \\
  Sridhar, S., 1987, J. Astrophys. Astron. 8, 257 \\
  Syer, D., 1995, MNRAS, 276, 1009 \\
  Syer, D., \& Tremaine, S. 1996, MNRAS, 282, 223 \\
  Tremaine, S., H\'{e}non, M., \& Lynden-Bell, D. 1986, MNRAS, 219, 285 \\
  Vine, S., \& Sigurdsson, S. 1998, MNRAS, 295, 475 \\
  Weinberg, M. D., 1993, ApJ, 410, 543 \\
\end{document}